\documentstyle[11pt,newpasp,twoside,epsf]{article}
\begin{document}
\title{The Two Micron All-Sky Survey: Removing the Infrared Foreground}
\author{John E. Gizis}
\affil{Infrared Processing and Analysis Center, 
California Institute of Technology, 
Pasadena, CA 91125 USA}
\author{Michael F. Skrutskie}
\affil{Dept of Astronomy, University of
Massachusetts, Amherst, MA 01003 USA} 

\begin{abstract}
We introduce the properties of the Two Micron All-Sky Survey (2MASS) 
survey for IAU Symposium 204.  2MASS is a 
near-infrared survey of the entire sky characterized
by high reliability and completeness.  Catalogs and images
for 47\% of the sky are now available online.  This data
release has been used by Wright (2000) and
Cambr\'esy et al. (2000) to subtract the stellar foreground at
1.25 and 2.2 microns from COBE DIRBE data, revealing the
cosmological near-infrared background.  
\end{abstract}

\section{Introduction}

The Two Micron All-Sky Survey (2MASS) is an ongoing survey to 
map the entire sky at near-infrared wavelengths.  Motivated
by the particular application of 2MASS to the
determination of the extragalactic near-infrared background, 
our goal in this paper is to review the status and quality of 2MASS.
The 2MASS dataset has many applications, and it is our hope that this 
introduction to the data will motivate other investigators to
utilize the 47\% of the sky presently publically available.
The 2MASS catalogs and images are available online at 
http://www.ipac.caltech.edu/2mass/ 
and are fully described by Cutri et al. (2000).  

\section{Overview of 2MASS}

\subsection{Survey Characteristics}

The primary goal of 2MASS is to produce a {\it uniform} and
{\it unbiased} near-infrared survey of the entire sky characterized
by high reliability and completeness.  It is thus a modern
version of the classic Palomar sky survey (POSS) -- but with digital,
fully calibrated data in a new wavelength regime (Fig. 1).   2MASS is 
 a joint project of the University of Massachusetts and the Infrared
Processing and Analysis Center (IPAC)/California Institute of Technology, 
funded by the National Aeronautics and Space Administration and 
the National Science Foundation.  The primary task of U. Mass 
has been to fabricate and operate two 1.3-meter telescopes
(one at Mt. Hopkins, Arizona, and the other at CTIO, Chile) equipped
with multi-band infrared cameras.  IPAC has developed and maintained
a data pipeline capable of routine processing of 20 GBy of raw
data per day into source lists and images.  The results
must be distributed to the scientific community in a timely fashion.

\begin{figure}
\plotone{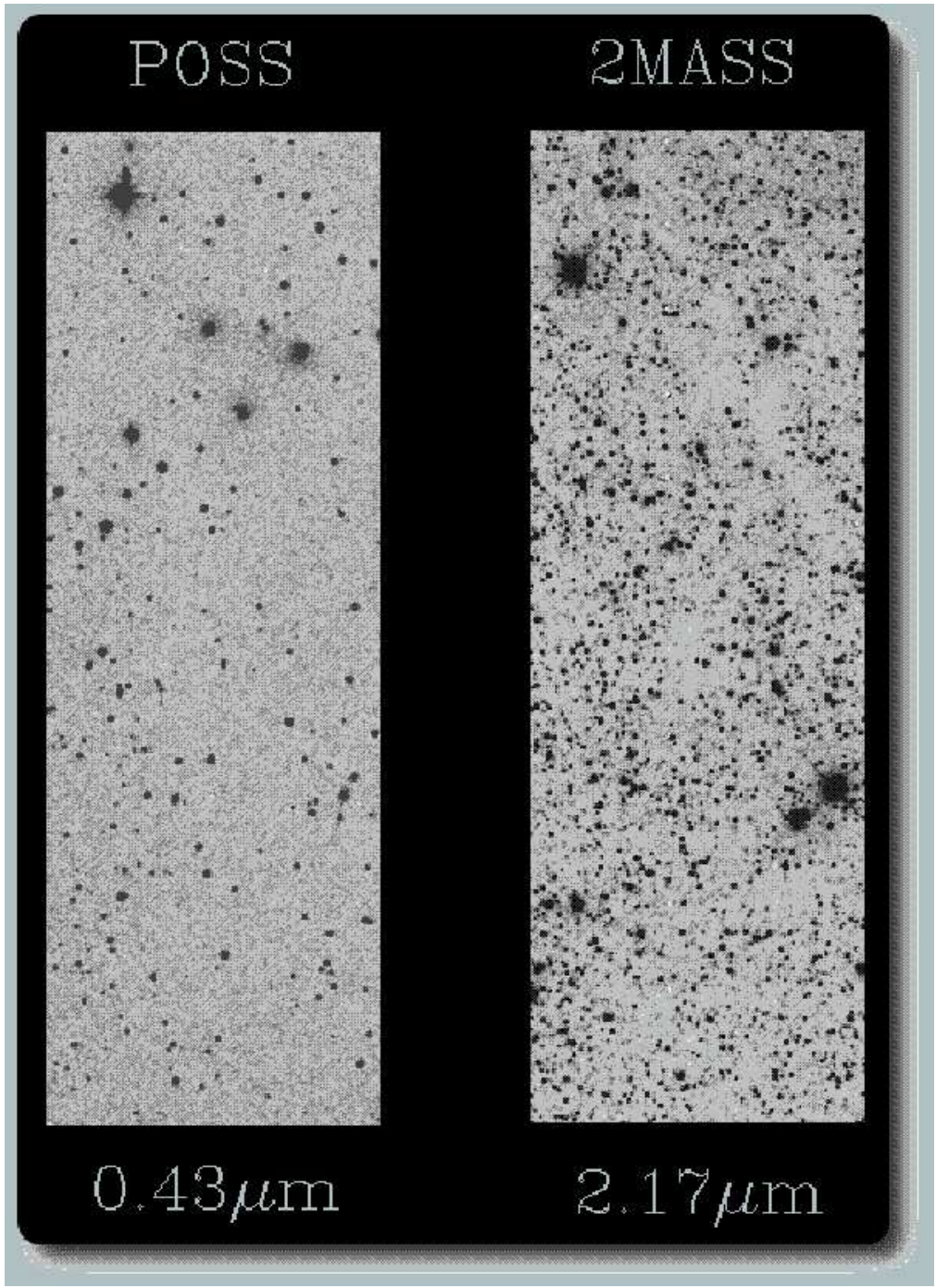}
\caption{POSS optical observations compared to 2MASS in the Galactic Plane}
\end{figure}

Some of the characteristics of 2MASS are summarized in Table 1.
The cameras use beam-splitters to obtain J, H, and K$_s$ simultaneously.
This is an important advantage of 2MASS, since there is no possibility
of producing anamolous colors due to variable or transient objects.
If an object found in one band is not detected in other bands, 
upper limits on its brightness are measured.    

While uniformity of the survey is a primary goal, varying
observing conditions (background and seeing) have effects on the 
achieved sensitivity.
Observations that do not meet the sensitivity requirements in
Table 1 are rescanned.  The K$_s$ backgrounds are driven by
thermal emission --- as a result, winter observations at
Mt. Hopkins are as much as 0.7 magnitudes deeper than the requirements,
imprinting a systematic effect on the sky.  
The H backgrounds are dominated by airglow which is 
highly variable and not very seasonal dependent.    

\begin{table}
\caption{2MASS Characteristics}
\begin{tabular}{ll}
Arrays&256$\times$256 NICMOS3 (HgCdTe)\\
Wavebands& J, H, and K$_s$ (2.00 -- 2.32$\mu$m)\\
Telescopes& 1.3-meter Equatorial Cassegrain\\
Pixel Size&2.0$^{\prime\prime}$\\
Integration time & 6 $\times$ 1.3s/frame = {7.8s} total\\
Sensitivity (10$\sigma$)&15.8, 15.1, 14.3 mag  for J, H, and K$_s$\\
Photometric Accuracy& 5\% for bright sources (3\% in practice)\\
Photometric Uniformity& 4\% over the sky  (1\% in practice)\\
Positional Accuracy&0.5$^{\prime\prime}$ (0.2$^{\prime\prime}$ in practice)\\
Completeness / Reliability&0.99 / 0.9995
\end{tabular}
\end{table}

The unit of the 2MASS survey is the 6$^\circ$ $\times$ 8.5$^\prime$ tile.
Each tile consists of 273 overlapping exposures (Figure 2).  As a result,
each star is imaged 6 times and total exposure time is 
1.3 s/frame $\times$ 6 = 7.8 seconds.  The sky is divided into
$\sim 60000$ tiles, which overlap slightly in right ascension.  
The 6 samples allow bad pixels and cosmic rays to be eliminated, and
also 1$^{\prime\prime}$
pixels to be generated for the final data products from the
2.0$^{\prime\prime}$ camera pixels.  

\begin{figure}
\plotfiddle{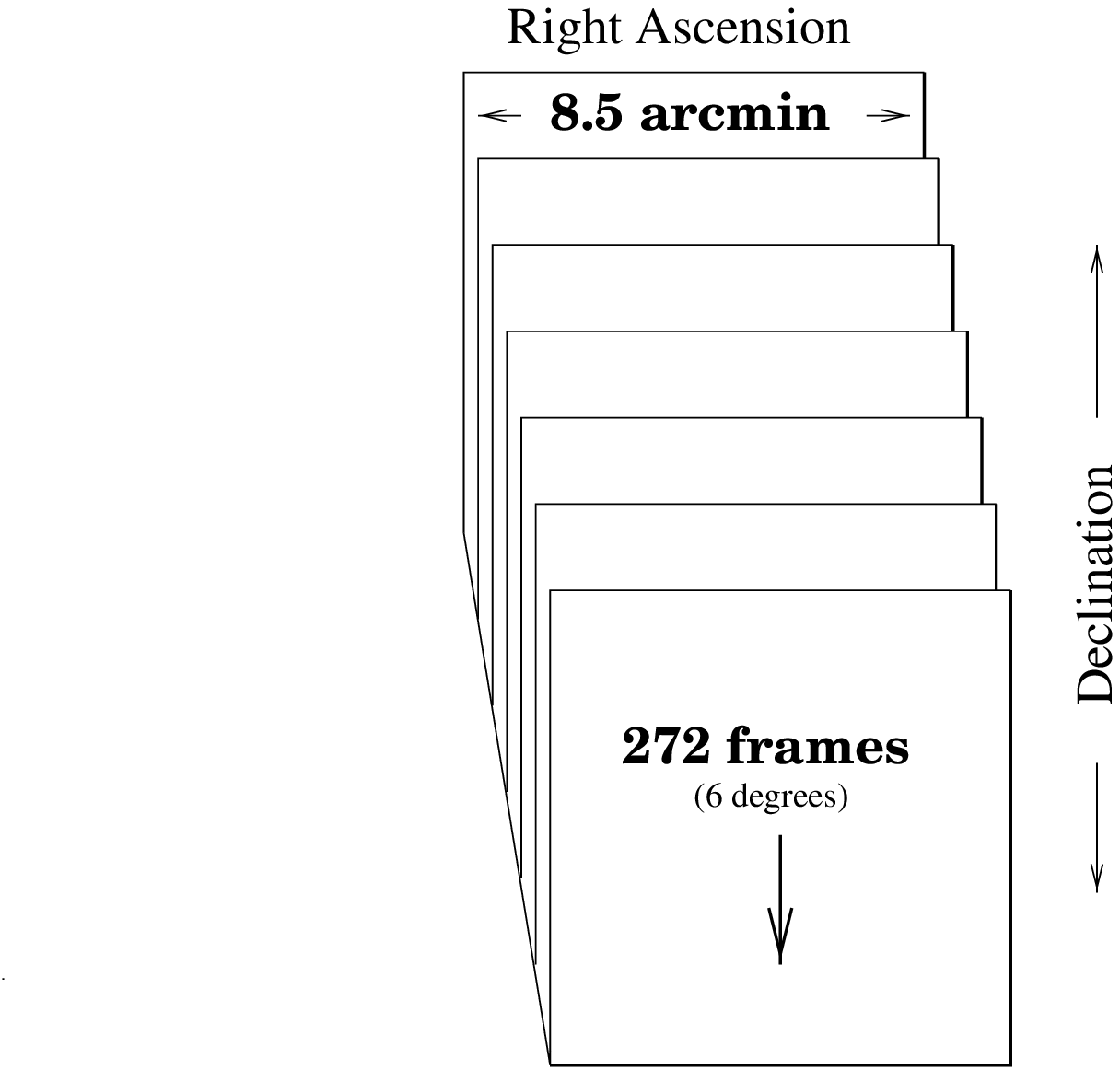}{4in}{0}{100}{100}{-250}{50}
\caption{2MASS Scanning Method:  Each point on the sky is imaged 6 times.}
\end{figure}

\begin{figure}
\plotone{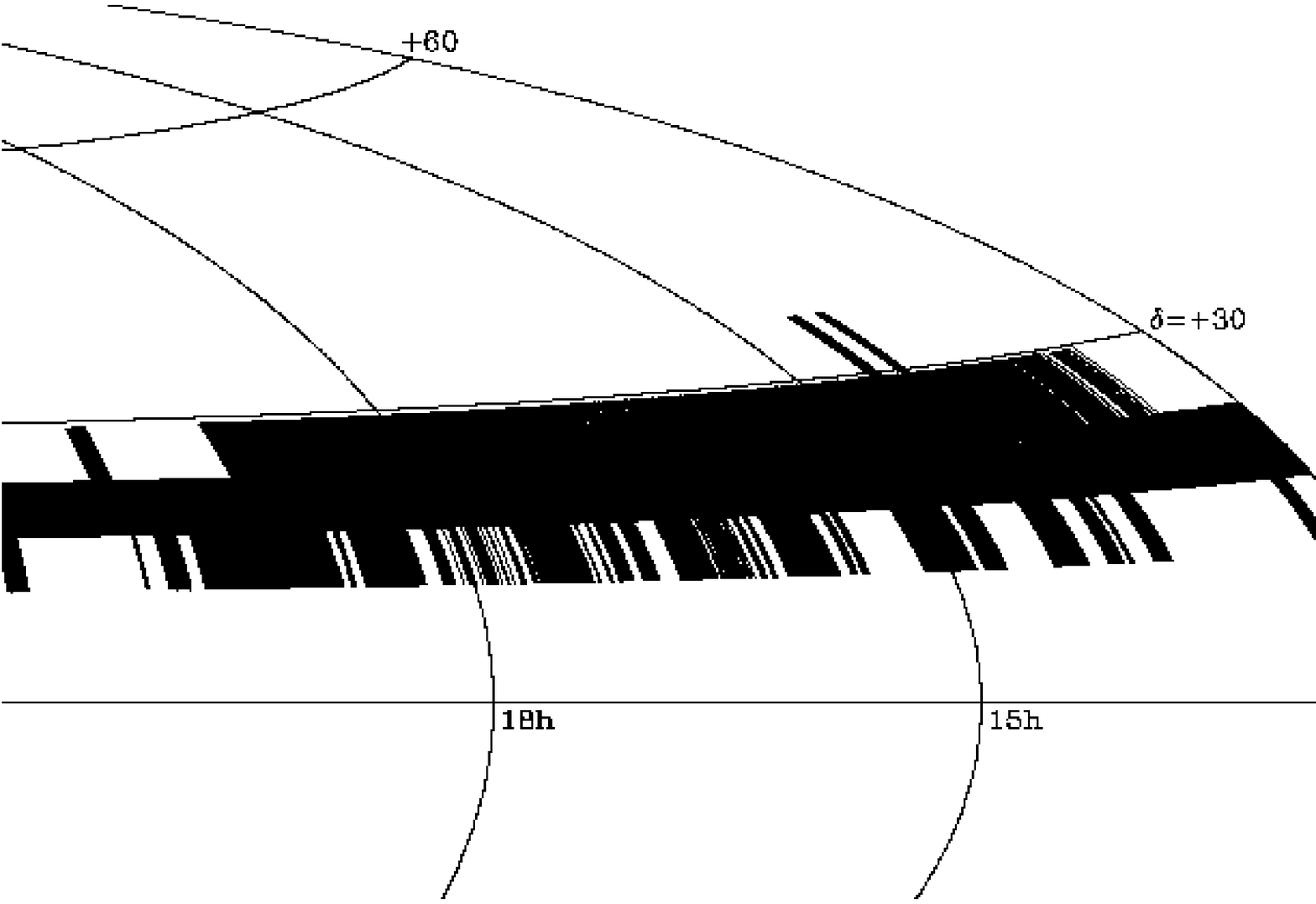}
\caption{2MASS Scanning Strategy: The sky is divided into overlapping tiles.}
\end{figure}

\subsection{Data Products}

The 2MASS data products consist of catalogs of sources and 
an atlas of images.  The {\it Point Source Catalog} (PSC) includes
$\sim 300$ million sources with astrometric and photometric
information.  Sources down to signal-to-noise ratios of 7
are included in the PSC.  The {\it Extended Source Catalog} (XSC)
includes $\sim 2$ million objects.  Jarrett et al. (2000a) 
give a full description of the extended source processing. 
The extended sources are identified on the basis of morphology,
and are mostly galaxies ---  the software has been designed to 
exclude double stars from the XSC.  Because galaxies are typically
redder (J-K$_s > 1$) than stars, a G-score is also available
using color information to improve the reliability of the galaxy catalog.
``Snapshot'' images of all extended sources are available online. 
It should be noted that the PSC includes many galaxies too
faint to be resolved --- most can be identified by their 2MASS
or 2MASS+optical colors.  

The 2MASS Digital Atlas of the sky consists of 1.8 million
calibrated 512x1024 images in the three passbands (11 TBy).  
Presently, the Quicklook atlas is available --- these images
have been lossy-compressed.  They are useful for finding charts
or visually verifying a source, but cannot be used for photometry.
The full, uncompressed images in FITS format are expected to be
available online soon.

The data products (presently 47\% of the sky) are available online
through the Infrared Science Archive (IRSA).  IRSA offers the
capability to efficiently query the catalog using SQL commands.
Source lists may be uploaded to IRSA for positional searches.  
DVD versions of the catalog releases are also available. 

\begin{figure}
\plotone{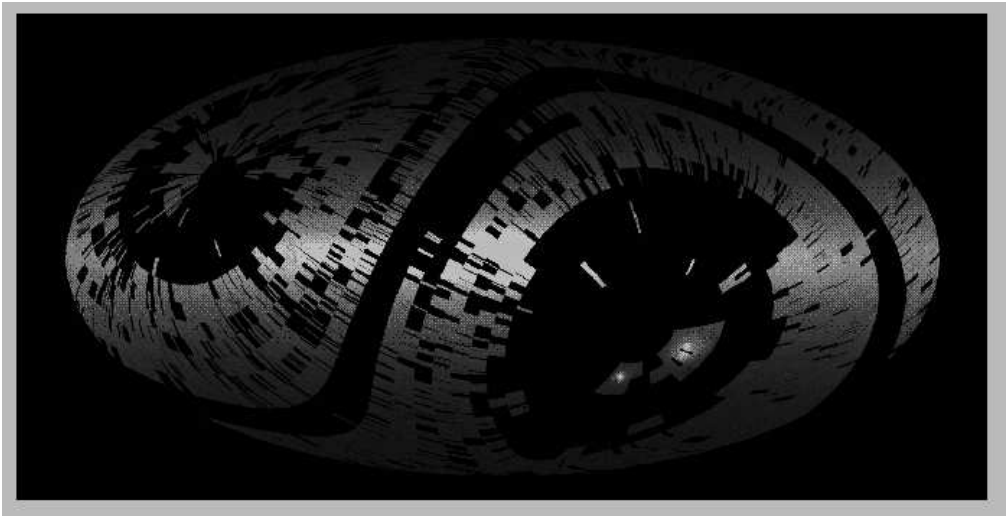}
\caption{Star Counts for the 2MASS Second Incremental Release.}
\end{figure}

\subsection{Photometric Quality}

Because it is so critical to so many applications --- particularly
the extragalactic background --- the photometric quality of 
2MASS deserves additional discussion.  Calibration is 
obtained once an hour by six repeated one-degree survey strips
(providing 36 observations of each star).  The strips are
centered on NICMOS calibrators (Persson et al. 1998).  Although
initially only the single NICMOS calibrators were available,
using the first year and a half of 2MASS data
Nikolaev et al. (2000) identified and calibrated $\sim 50$
secondary stars per field.  Their global analysis demonstrates
that the photometric system is uniform around the sky and in
time at the 1\% level.  There is no evidence for color
terms between the two hemispheres.  The nightly zeropoints
are determined to better than 1\% -- an example is shown
in Fig. 4.  Furthermore, the repeatability of {\it all} the 
high signal-to-noise stars in the strips are examined
to verify that conditions are photometric.  The RMS statistics
of the 6 repeats (Fig. 5) are computed to monitor the
sensitivity of the system.  The extensive dataset from these
observations allow us to quantify the sensitivity in terms
of the background and seeing, allowing the quality of the
science scans to be assessed.  Clouds between calibration
observations can
be identified by comparing the photometry in science scan
overlap regions and by their high and variable K$_s$ emission.

\begin{figure}
\plotone{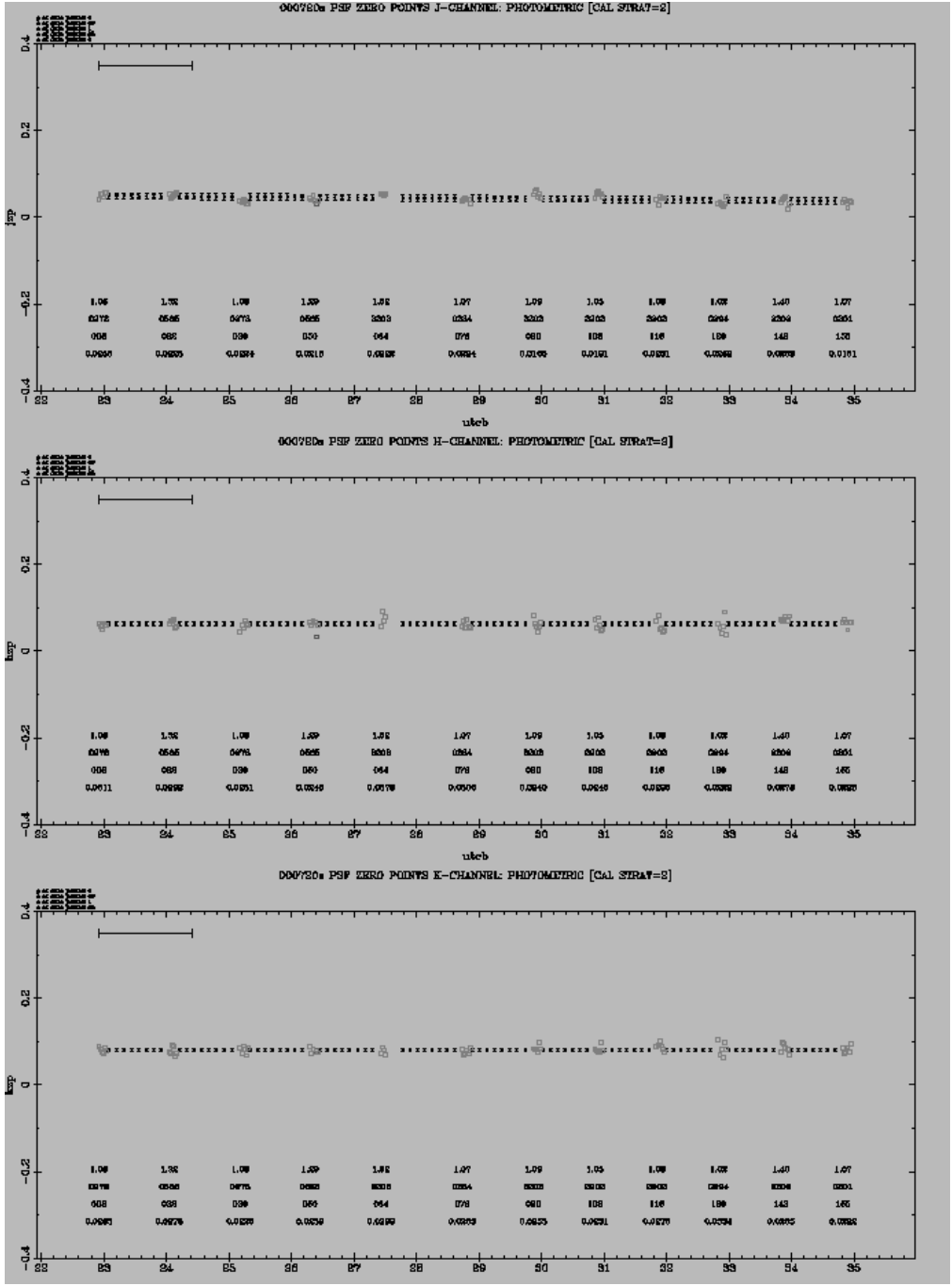}
\caption{Zero-point solution for 20 July 2000 CTIO observations.}
\end{figure}

\begin{figure}
\plotone{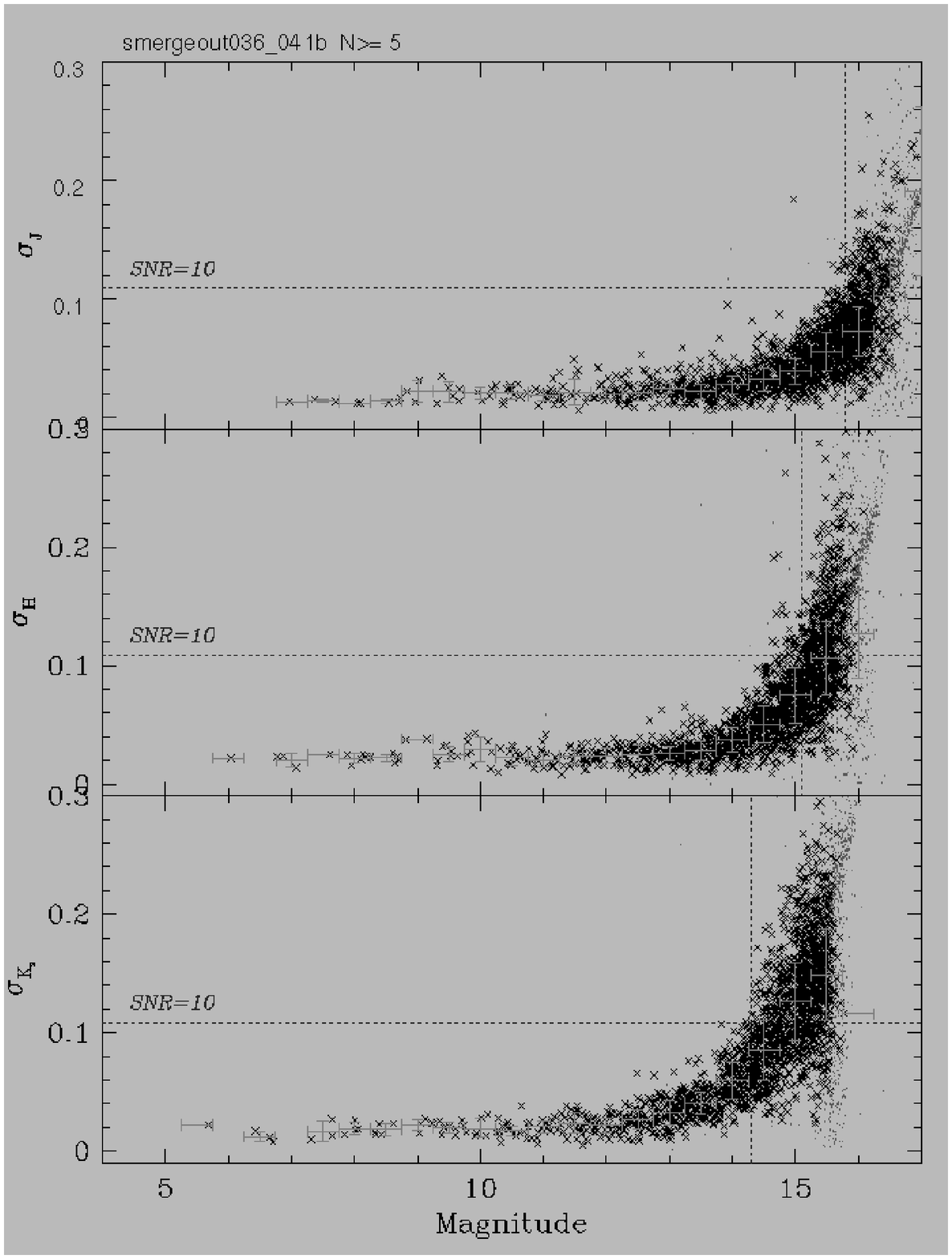}
\caption{Repeat statistics for one calibrator set on 20 July 2000.}
\end{figure}

\section{Science Applications}

The near-infrared offers two special advantages over optical observations.
First, interstellar dust is more transparent in the IR ($A_K \approx 0.1 A_V$).
As a result, the ``Zone of Avoidance'' cased by the Milky Way is reduced.
Furthermore, internal extinction, such as star formation
regions and other galaxies, is minimized.  Second, the near-infrared
is more sensitive to ``cool'' objects.  This enables the study of
IR-selected populations underrepresented in optically-selected
samples.  For galaxy studies, the older stellar population
emits a larger fraction of IR light, giving a better handle
on the true mass distribution.  

As a taste of the diverse uses of 2MASS data, we briefly
mention some recent studies.  Within the solar system,
Sykes et al. (2000) show that 2MASS data can classify 
asteroids on the basis of their composition.  The census
of the local solar neighborhood is being completed by
the discovery of very-late M dwarfs (Gizis et al. 2000),
L dwarfs (Kirkpatrick et al. 1999), and T Dwarfs (Burgasser
et al. 2000) ---  L and T dwarfs are new spectral classes
($T_{eff} < 2200$K)    
of brown dwarfs.  Star counts constrain Galactic structure
(Ojha, this volume) and show the existence of flaring and warping
of the outer Galactic disk (Alard 2000).  

2MASS is also contributing to extragalactic astronomy.
Still within the local group, the structure of the Large 
Magellenic Cloud --- and
the interpretation of MACHO microlensing results ---
is discussed by Weinberg \& Nikolaev (2000).  Large-scale
structure behind the Zone of Avoidance is revealed
by Jarrett et al. (2000b).  Infrared selection has
uncovered a population of very red quasars and AGN (Nelson et al. 1999).   
Remarkably, 2MASS also is relevant for studies of the
cosmological background, discussed in the next section.

\section{The Near-IR Foreground and Extragalactic Background}

The use of 2MASS for determining the cosmological background
has been pioneered by Wright (2000), whom we follow in 
this discussion and which the interested reader is urged
to consult. The DIRBE experiment
on COBE (see Hauser, this volume) provides calibrated observations
of the entire sky at 1.25 and 2.2 microns, corresponding
to the 2MASS J and K$_s$ filters.  The DIRBE beam size is $0.7\deg \times 
0.7\deg$ --- thus, many stars are included in each DIRBE beam.  
Arendt et al. (1998) found that the uncertainties and systematic
errors in galactic star
count models are large enough that a detection of the infrared
background at 1.25 and 2.2 microns was not possible.  They report
that the uncertainty due to stars is 15 nW m$^{-2}$sr$^{-1}$ 
at 1.25 microns and 10 nW m$^{-2}$sr$^{-1}$ at 2.2 microns ---
to be compared with an estimated uncertainty of 15 and 6 respectively 
in the Zodaical light model.  Since the cosmological
background is approximately this magnitude --- and since 
Arendt et al. find systematic residuals as a function of
galactic latitude --- a more direct approach is needed.

Wright (2000) has approached this problem by using the 2MASS 
data release to directly estimate the contributions of
stars to the observed DIRBE fluxes.  (Gorjian, Wright, \&
Chary 2000 used their own 2.2 and 3.5 micron observations to 
analyze a single dark spot.)  Selecting DIRBE dark regions
minimizes the contribution of zodiacal light.  Wright found
four DIRBE dark regions  with complete 2MASS coverage
in the data release --- even in these pixels, 
zodiacal light contributed
$\sim 75\%$ of the observed DIRBE 2.2 micron signal.  
A further $\sim 15\%$
is contributed by foreground stars.  The remaining signal 
is due to the desired extragalactic background.  (Cambr\'esy
et al. (2000) estimate that the 2MASS-resolved galaxies contribute only 
5\% of the light.)  Wright has shown that the 2MASS data
allows the stellar contribution to be subtracted 
from the DIRBE data on a pixel-by-pixel basis --- there is
an excellent correlation between the 2MASS integrated star
brightness and DIRBE flux.  This allows
a detection of the extragalactic background light and for
strict limits to be placed on fluctuations in the background.   
Wright (2000) uses the 'no-zodi principle' zodiacal light models 
of Wright (1998).  

The star-subtraction approach has been extended by 
Cambr\'esy et al. (2000),
who use the public 2MASS PSC  to subtract the
stellar contribution over a large area of the sky.  This
allows them to confirm the {\it isotropy} of the 
putative extragalactic background --- a sign that the
background is truly cosmological.  2MASS's determination of the
stellar contribution is precise enough that the
zodiacal modelling dominates the uncertainty, although 
the contribution of very faint stars and the flux calibration
make small contributions (see the
discussions throughout this volume).
In particular, use of the Kelsall et al. (1998) zodiacal light model
makes Cambr\'esy et al.'s estimated cosmological background 
higher than Wright's (2000) estimates.  Wright estimates
that 94\% of the uncertainty in the cosmological light 
is due to the zodiacal light model at both 1.25 micron
and 2.2 microns.  2MASS has thus allowed the stellar
uncertainty to be reduced from being the dominant term
(Arendt et al. 1998) to being a negligible one.  
For the actual results and interpretation, see Wright (2000) and 
Cambr\'esy et al. (2000).

\section{Summary}

Nearly half the sky is now publically available.  At the
time of this writing (11 October 2000), 99.8\% of the sky
has been observed with catalog-quality data.   The present
data release to the astronomical community has enabled 
studies from the solar system to cosmology
even before the final processing of the full 2MASS dataset.

\end{document}